\documentclass[lettersize,journal]{IEEEtran}
\usepackage{cite}
\usepackage{amsmath,amssymb,amsfonts}
\usepackage{algorithmic}
\usepackage{graphicx,color}
\usepackage{textcomp}
\usepackage{cite}
\usepackage{hyperref}
\usepackage{amsmath,amsfonts}
\usepackage{color}

\usepackage{url}
\usepackage{graphicx}
\usepackage{float}
\usepackage[table,xcdraw]{xcolor}
\usepackage{multirow}
\usepackage{siunitx}
\usepackage{chemformula}
\usepackage[caption=false,font=normalsize,labelfont=sf,textfont=sf,subrefformat=parens,labelformat=parens]{subfig}
\usepackage{verbatim}
\usepackage{stfloats}
\usepackage{textcomp}
\usepackage{nomencl}
\usepackage[acronym]{glossaries}
\usepackage{tabularx}
\usepackage{color,soul}
\usepackage{array}
\usepackage{pifont}% http://ctan.org/pkg/pifont
\usepackage[normalem]{ulem}
\newcommand{\cmark}{\ding{51}}%
\newcommand{\xmark}{\ding{55}}%
\def\BibTeX{{\rm B\kern-.05em{\sc i\kern-.025em b}\kern-.08em
    T\kern-.1667em\lower.7ex\hbox{E}\kern-.125emX}}
\begin{document}

\title{Reconfigurable Intelligent Surface Deployments for Mars: Communication and Localization Across Diverse Terrains}

\author{Enes Koktas,~\IEEEmembership{Student Member,~IEEE}, Recep A. Tasci,~\IEEEmembership{Student Member,~IEEE}, Ibrahim  Yildirim,~\IEEEmembership{Graduate Student Member,~IEEE} and Ertugrul Basar,~\IEEEmembership{Fellow,~IEEE} 
		% <-this % stops a space
		\thanks{E. Koktas, Recep A. Tasci, I. Yildirim, and E. Basar are with CoreLab, Department of Electrical and Electronics Engineering, Ko\c{c} University, Istanbul, Turkey (e-mail: enes.koektas@kit.edu, recep.tasci@medipol.edu.tr, ibrahimyildirim19@ku.edu.tr, ebasar@ku.edu.tr).}
        }
\maketitle
\begin{abstract}
Space exploration has witnessed a steady increase since the 1960s, with Mars playing a significant role in our quest for further knowledge. As the ambition to colonize Mars becomes a reality through the collaboration of private companies and space agencies, the need for reliable and robust communication infrastructures in the Martian environment becomes paramount. In this context, reconfigurable intelligent surface (RIS)-empowered communication emerges as a promising technology to enhance the coverage due to lack of multipath components in line-of-sight (LOS) dominated Martian environments. By considering various Martian scenarios such as canyons, craters, mountains, and plateaus, this article explores of the potential of RISs in increasing the coverage in Martian environments. The article also provides an overview of RIS-assisted localization in both LOS and non-line-of-sight (NLOS) scenarios, presenting a general framework for accurate user angle estimation in challenging Martian conditions. The findings and presented framework of this article provide a promising research direction for integrating RISs in deep space communication as well as paving the way for future improvements in interplanetary communication networks.
\end{abstract}


\section{INTRODUCTION}
\IEEEPARstart{M}{ars} has remained the central focus of interplanetary exploration for humankind for more than half a century. Engineers from across the world have dedicated immense effort to Mars missions, driven by the prospect of gaining vital insights into our planet's future and better preparing ourselves for existential challenges. In addition, these missions hold the promise of establishing the first outer space habitat, paving the way for further exploration of other planets. The Martian environment poses unique challenges compared to Earth, including sparse transceiver deployment, negligible multipath propagation due to the surface’s low reflection coefficient, frequent dust storms, and limited power resources. Overcoming these obstacles, particularly the need for connectivity beyond line of sight, calls for autonomous self sustaining modules capable of ensuring uninterrupted coverage throughout Mars missions. Conventional terrestrial base stations that underpin Earth’s cellular networks rely on robust power grids, maintenance crews, and transportation infrastructure; these resources are unavailable on Mars and render such installations impractical. Base stations (BSs), which are the main provider of cellular communication on Earth, appear as the first solution to be called by. However, we believe that deploying active BSs on Mars is neither a practical nor an economical approach. On Earth, BS installations lean on existing power grids, robust maintenance crews, and a supportive transportation network; such infrastructure is absent on the Red Planet. On the other hand, we assert that reconfigurable intelligent surface (RIS)-assisted systems hold tremendous potential for Mars missions. RIS modules can be made lighter, simpler, cheaper and do not require continuous power supply compared to traditional  BSs. Rovers can transport or place modular and solar-powered RIS panels easily, whereas BS deployments require more elaborate setups. In future Mars missions, where a rover is exploring a canyon hundreds of kilometers from any installed Martian base, a small, low-power RIS panel can be quickly placed at the canyon edge to create a reliable link.
    
    RIS-empowered wireless communication has gained significant interest as an emerging technology for sixth-generation (6G) networks. Specifically, it offers a low-cost and energy-efficient solution to achieve enhanced spectral efficiency. By dynamically adjusting signal propagation using passive reflecting elements of an RIS, wireless coverage and signal quality can be enhanced while maintaining cost-effectiveness and low hardware complexity.
    The concept of RIS involves deploying tiny and adaptive materials on various structures, including building walls and ceilings. While RISs are not capable of establishing communication links independent from base stations, they offer a significant contribution to augmenting the performance of existing infrastructure without imposing notable processing costs. 
    RISs might have potential applications in Mars communication, where direct links between transmitters and receivers may be disrupted by challenging terrains and the signal may completely degrade after first reflection. Furthermore, RISs can be utilized to increase the throughput of orbiter-rover links similar to the integration of RISs with low Earth orbit (LEO) satellite networks to address frequency constraints and enable low-complexity system design for Earth scenarios as investigated in \cite{Tian_2022_Enabling_NLoS_LEO_Satellite}. In \cite{10993923}, the authors employed the ns-3 simulator to evaluate commercial LoRaWAN IoT network performance on Mars, explicitly modeling free-space path loss and attenuation caused by Martian dust storms as factors of link distance, packet size, and traffic load. Their findings indicated that the throughput on Mars drops more rapidly with distance compared to Earth. Nonetheless, communication over fairly long distances, approximately 2000 meters, can still be accomplished by decreasing the offered traffic. Furthermore, they noted that, at the frequencies utilized by LoRa, dust storms do not substantially impact the system's performance.
    
\begin{figure*}[t]
        % \centering\includegraphics[width=1\linewidth]{figures/systemmodel1v2.png}
        \centering\includegraphics[width=1\linewidth]{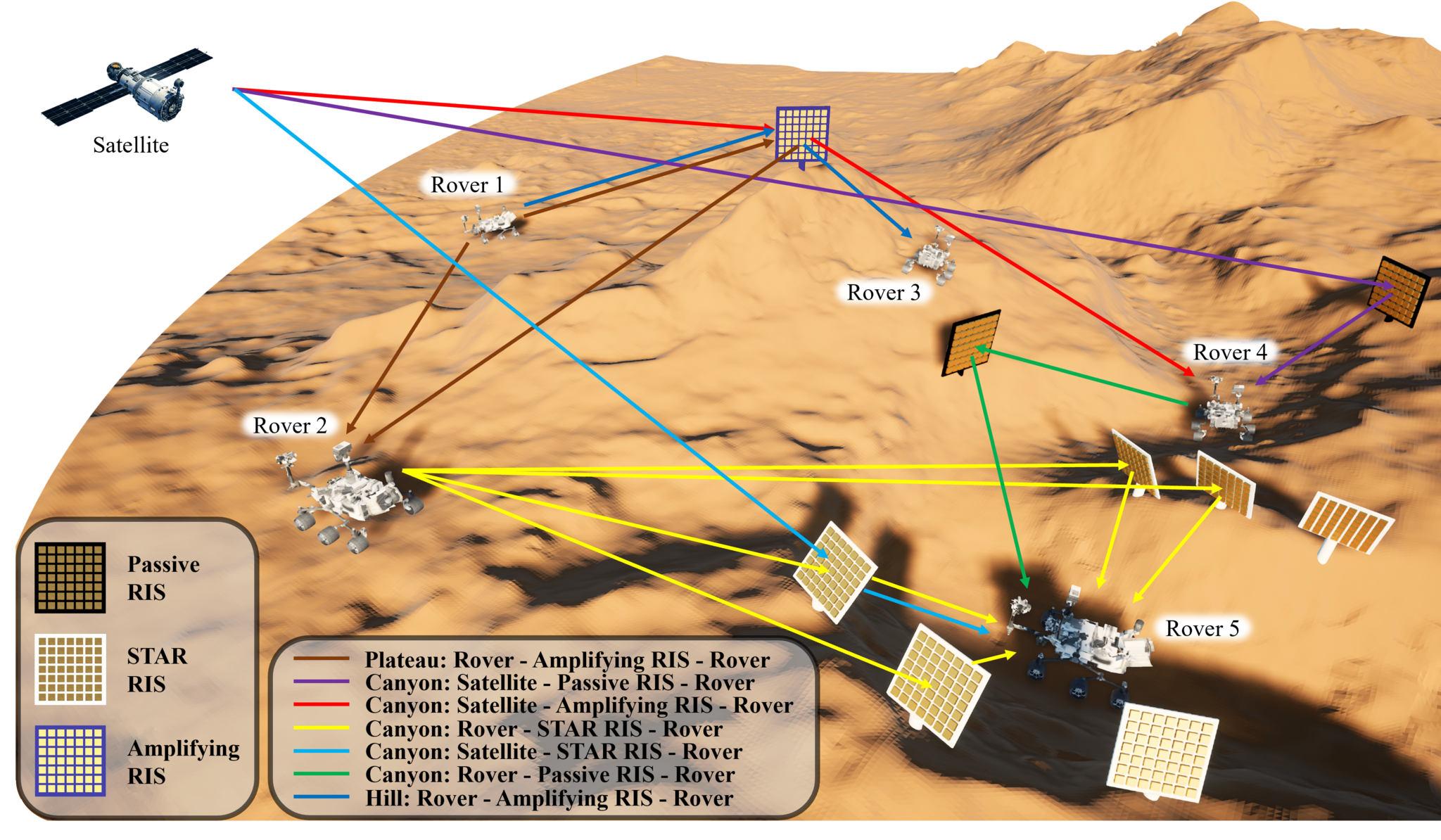}
        \vspace{10pt}
        \caption{RIS-assisted Martian communication network on Jezero Crater.}
        \label{fig:ntwmodel}%\vspace*{-0.3cm}
    \end{figure*}

    In this article, we conducted a comprehensive examination of the performances of various RIS types in diverse Martian environments, encompassing crater, canyon, mountain, and plateau landforms as shown in Fig. \ref{fig:ntwmodel}, which is created using 3D model of Jezero crater, landing site of Perseverance rover. However, the 3D model is scaled for illustration purposes. To ensure the most efficient utilization of RISs in these challenging terrains, we also give attention to localization-based implementations, as they reduce the additional power consumption associated with channel state information (CSI) acquisition. In general, our discussions and analysis focus on numerous aspects, including coverage enhancement, data throughput improvement, energy efficiency, and system complexity, with the ultimate goal of advancing the reliability and robustness of Mars communication systems. Furthermore, we believe that the insights and frameworks established in this article not only  deliver a promising research direction for integrating the RIS technology into Martian communication but also lay the foundation for extending our approaches to communication on different planets in the future. %Furthermore, we believe that the insights and frameworks established in this article could be applicable to future exploration missions on other celestial bodies as well. 
     As humanity's ambitions reach beyond Mars, the findings obtained from this study can serve as a valuable blueprint for optimizing communication networks on other celestial bodies, thus encouraging interplanetary exploration and communication advancements.

    The rest of the article is summarized as follows: Section \ref{sec:overview_mars} provides a brief overview of Mars communication, including its unique characteristics and challenges. A general overview of RIS types and RIS-assisted localization is given in Section \ref{sec:overview_ris_types} and \ref{sec:ris_ass_loc}, respectively. Section \ref{sec:ris_in_mars} discusses the deployment of RIS-empowered communication and localization in Mars communication concerning various Martian landforms. Finally, Section \ref{sec:Conclusions} concludes the article.

    \section{An Overview of Mars Communication Challenges}
    \label{sec:overview_mars}
    Intra-Mars communication presents several challenges, particularly pertaining to radio propagation. Key obstacles encompass severe signal attenuation, signal delay, Doppler shift, atmospheric interference, and power constraints. Radio signals traversing the Martian atmosphere encounter signal attenuation due to factors such as atmospheric composition, density, and dust storms, necessitating the consideration of these variables for reliable communication. Power constraints in the Martian environment impose energy-efficient transmission techniques, power management strategies, and hardware optimization to ensure reliable communication with minimal power consumption. Lastly,  it is of utmost importance to ensure high reliability, with the need for communication systems involving multiple ground stations and orbiting satellites to provide continuous coverage and contingency measures in the face of failures or anomalies. Continuous monitoring and analysis of communication links are imperative to optimize performance and surmount the inherent limitations of radio propagation on Mars.
        \subsection{Propagation characteristics and challenges}
        Mars has a significantly different atmospheric composition compared to Earth, with a thin atmosphere primarily composed of carbon dioxide.
        The attenuation of radio signals in the Martian atmosphere is influenced by molecular absorption and scattering. The low-density atmosphere and the presence of carbon dioxide can cause significant signal loss, particularly at higher frequencies. 
        Secondly, the extreme temperature variations on Mars further impact signal propagation. The diurnal temperature variations can lead to fluctuations in the refractive index, causing signal bending, dispersion, and multipath effects. 
        However, the studies conducted by the National Aeronautics and Space Administration (NASA) show that these atmospheric effects can be negligible at sub-6 GHz frequencies \cite{koktas2022communications}. We anticipate that the performance of LOS communication links on the Martian surface will be close to free-space links.
        
        Beyond the effects of the Martian atmosphere, the Martian surface itself presents challenges to signal propagation. The irregular terrain, including hills, valleys, and craters, can cause signal blockage, shadowing, and diffraction effects. These terrain-induced obstructions can significantly degrade signal strength and introduce variations in received signal power, particularly in non-line-of-sight (NLOS) scenarios.

        The electrical properties of the Martian surface differ significantly from those of the Earth's surface, resulting in distinct reflection coefficients for the two planets. Earth's surface exhibits an electrical conductivity on the order of $10^{-4}$, whereas Mars has a considerably lower electrical conductivity, ranging from $10^{-12}$ to $10^{-7}$ and relative permittivity in the range $2.5$–$9$, producing weaker surface reflections and fewer natural multipath components than on Earth\cite{SIMOES200830}. On the other hand, our knowledge of the reflection coefficient of the Martian surface is limited due to limited experimental data. However, because of the discrepancy in electrical conductivities between the two planets, it is anticipated that Mars' reflection coefficient to be lower than Earth's. Flat terrains therefore approach free-space behavior with path-loss exponents near 2 (e.g., 2.12 in the flat sector of Gale crater), while rocky sectors modestly increase the exponent to about $2.37$. Reported root mean square (RMS) delay spreads on Mars are typically well below outdoor Earth values \cite{koktas2022communications}. These contrasts motivate RIS on Mars as a means to synthesize controllable reflecting paths where the natural environment offers limited multipath support. Consequently, the utilization of multipath components for communication purposes is more challenging in the Martian environment. Due to the difficulty in leveraging multipath components, the presence of a LOS link between the transmitter and receiver becomes crucial for establishing reliable communication on Mars. RISs are employed to establish these LOS links between the transmitter and receiver, particularly in scenarios where NLOS conditions exist, and offer a promising solution to overcome the challenges imposed by obstructed paths and signal blockages in the Martian environment.

        \subsection{"Advantages and Considerations in Deploying RIS Technology on Mars"}

        RIS stands out due to its low repair and maintenance requirements, which distinguishes it from the more demanding requirements of typical relay base stations \cite{9998527}. However, the prospect of RIS overcoming harsh Martian conditions remains an unknown variable. Despite this uncertainty, deploying RIS technology to Mars has the advantage of being a low-cost operation. In contrast, the total costs of traditional relay base stations, including transportation, deployment, and ongoing maintenance, are expected to substantially exceed the economic benefits of deploying RISs. It is important to emphasize that, unlike traditional infrastructure, RIS adoption has no specific costs, which contributes to its cost-effectiveness. In essence, we propose a practical approach adapted for a vehicle environment that leverages RISs without the need to build a whole communication infrastructure, demonstrating its effectiveness in addressing specific challenges.

    \section{General Overview of RIS Types}
    \label{sec:overview_ris_types}
\begin{figure*}[t]
        \centering\includegraphics[width=1\linewidth]{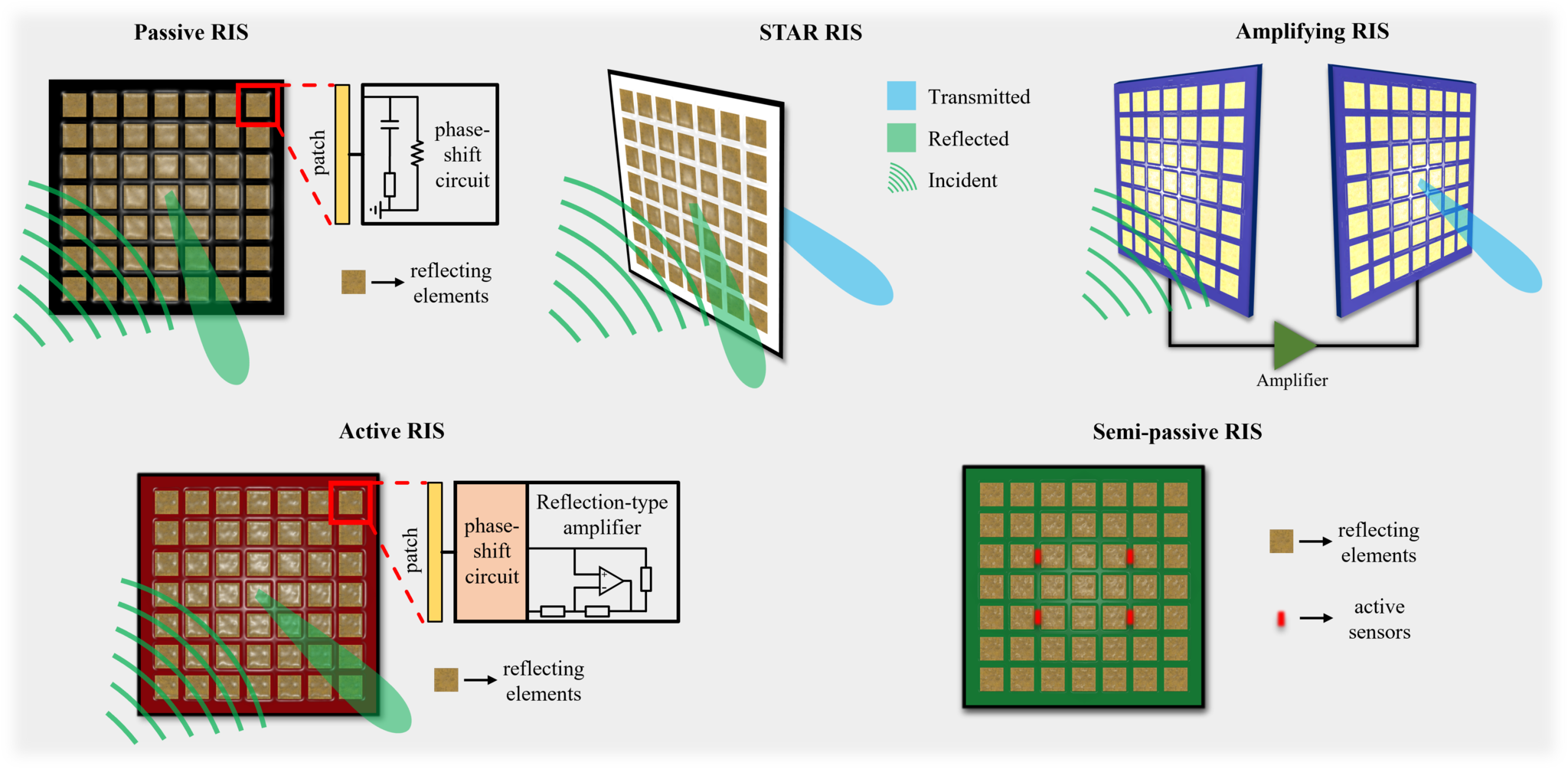}
            %\vspace{10pt}
            \caption{Visual representation of various RIS types.}
            \label{fig:RIStypes}%\vspace*{-0.3cm}
        \end{figure*}

    In this section, a concise overview is presented regarding various RIS types, including passive, semi-passive, active, amplifying, and simultaneously transmitting and reflecting (STAR)-RIS, each possessing specific advantages and disadvantages over different scenarios. These RIS types are illustrated in Fig. \ref{fig:RIStypes}.

        \subsection{Passive RIS}
        Passive RISs are both low-cost and energy-efficient due to the extensive use of passive reflecting elements in their hardware. The incident electromagnetic waves are intelligently redirected in the desired direction by the controller circuit, which skillfully controls the phase shifts on these passive components of the RIS \cite{ebasarwctris}, resulting in a considerable increase in the received signal-to-noise ratio (SNR). On the other hand, there is always the double path loss issue that leads to a drop in performance when the RIS is not placed near the communicating terminals. As a result, passive RISs can only serve as a supportive technology for communication systems, as they cannot overcome the double path loss problem. To address this challenge, several studies have explored the combination of active elements with RISs to mitigate the effects of double path loss. Some of these designs are briefly explained in Subsections \ref{sub_ActiveRIS} and \ref{sub_AmpRIS}. Channel estimation is another challenge for passive RISs, and semi-passive RISs, which are concisely mentioned in subsection \ref{sub_semipasRIS}, are used to overcome this problem.
        \subsection{Semi-Passive RIS}\label{sub_semipasRIS}
        Semi-passive RISs incorporate active sensors alongside the passive reflecting elements found in traditional passive RISs. These active sensors enable the acquisition of CSI within the RIS. Extensive research has been conducted to explore the capabilities and potential applications of these active sensors \cite{semipassiveRIS}.
        \subsection{Active RIS}\label{sub_ActiveRIS}
        Active RISs exhibit similarities to their passive counterparts in terms of their ability to reflect incident signals with adjustable phase shifts. However, active RISs distinguish themselves from passive RISs by possessing the capability to amplify the reflected signals. One way of doing this is the integration of active reflection-type amplifiers as a crucial component within each active RIS element \cite{9998527}. In addition to the signal amplification capability, active RISs suffer from the drawback of increased power consumption and cost, particularly when a large number of reflecting elements are employed.
        \subsection{Amplifying RIS}\label{sub_AmpRIS}
        Amplifying RIS has a simple hardware architecture that integrates a single, variable gain amplifier to enhance signal reflections \cite{9758764}. This design incorporates two passive RISs and a power amplifier positioned between them. One of the RISs serves as the receiver and signal combiner, while the other RIS functions as the transmitter for the amplified signal. Effective positioning of the RISs, such as positioning them back to back, eliminates electromagnetic leakage between the RISs, ensuring signal isolation and integrity. The design operates exclusively within the RF domain, resembling the waveguide-based approach as opposed to the full-duplex multi-antenna decode-and-forward (DF) relays, which engage in down-conversion and baseband processing. Consequently, our design exhibits greater similarity to a full-duplex, multi-antenna, and amplify-and-forward (AF) relay. Nevertheless, notable distinctions exist. Traditionally, relays employ linear processing techniques, such as maximum ratio combining (MRC), and implement optimization algorithms for power allocation. Additionally, they incorporate unwieldy phase-shifter networks for transmit beamforming, thereby exposing themselves to loopback self-interference. This particular design offers notable advantages in terms of energy efficiency and cost-effectiveness compared to relays and alternative active RIS designs, as it employs a single power amplifier. 
        \subsection{STAR-RIS}
        In contrast to traditional passive RISs, the surface elements of STAR-RISs are not only used for reflection but also for simultaneous transmission of a signal \cite{9437234}. This simultaneous transmission and reflection capability provides additional degrees of freedom and potential benefits in wireless communication systems. Recently, the authors of \cite{9935303} established that STAR-RISs possess the ability to concurrently reconfigure signal propagation from distinct sides. This phenomenon is observed when reflection and transmission are conducted simultaneously on both sides of the STAR-RIS, referred to as dual-sided STAR-RIS.

\section{An Overview of RIS-assisted Localization}
    \label{sec:ris_ass_loc}
    In this section, we provide a generic overview of RIS-assisted localization, which might shed light on their use in Martian environments.
    
    RIS-assisted localization has emerged as a promising solution for enhancing localization accuracy. 
    Two different approaches can be identified for estimation: (i) the first approach is a direct localization method used for asynchronous scenarios, and (ii) the second approach is a two-stage method that involves estimating the location and orientation based on a set of signal features, such as time-of-arrivals (TOAs), angle-of-arrivals (AOAs), and received signal strengths (RSSs) \cite{Elzanaty_2021_Rec}.
    Compared to instantaneous CSI, location information demonstrates a more gradual rate of change, enabling the design of efficient RIS reflection patterns that reduce the computational complexity at the base station (BS) and minimize control link interactions. Additionally, leveraging the users' location, the deployment of RISs can be optimized considering actual terrain constraints and user distribution, leading to further improvements in system performance. 
    %There has been growing interest in simplifying beamforming design using location information, primarily obtained from Global Positioning System (GPS), which introduces additional overhead due to the need for GPS modules on the user side. However, the limitation in computation capability of user devices necessitates reduced complexity. Therefore, the development of RIS-assisted position perception poses a significant challenge.
    
    The incorporation of RISs in wireless localization also enables localization at both the transmitter and receiver sides, leading to unprecedented control over the electromagnetic field. Geometric beams can be designed for location-based beamforming without the need for full-band CSI. Moreover, sophisticated spatial techniques can be employed to mitigate interference, while long-term location information enables the construction of radio environment maps, facilitating proactive resource allocation.
    It is worth noting that RIS-assisted localization is not limited to a specific number of users, making it suitable for scenarios with a potentially unlimited number of users. 
    Another area of interest is integrating sensing capabilities into the RIS network, allowing it to estimate the user's position more accurately. By modifying the reconfigurable elements of the RIS, a portion of the incident wave is coupled to a waveguide, enabling the RIS to sample the signal for deducing necessary information about the propagation environment. This approach reduces computational complexity and improves localization accuracy by avoiding the need to search all regions for fine estimation \cite{Alamzadeh_2021}.
    Furthermore, an RIS is integrated into the wireless localization system in \cite{Liu_2020}, focusing specifically on passive reflect beamforming to minimize localization errors. The authors established a 3-D model for the RIS-assisted wireless localization system. Then, the authors derived the Fisher information matrix and the Cramer-Rao lower bound, which provide insights into the accuracy of estimating the absolute position of the mobile station (MS). The results of the convergence analysis demonstrate that through the optimization of the reflect beamforming, it is possible to improve the accuracy of positioning, and achieve decimeter-level or even centimeterlevel positioning by employing the RIS with a large number of elements.

    Our treatment of localization aims to define how position information can drive RIS operation in Martian environments rather than to benchmark specific algorithms. The accuracy, latency, and robustness of a given method depend on anchor geometry, sensing modality, and terrain features, and these choices are mission-specific. In our framework, position estimates are used to select among precomputed RIS phase profiles or to trigger a reconfiguration when the long-term SNR target cannot be sustained. A full quantitative study of Mars-grade localization, including accuracy versus update-rate trade-offs and sensitivity to intermittent links, is an important next step and is left for future work.
    
    In light of this discussion, we now delve into the application of RISs and localization techniques in the Martian environment.  
    
    \section{RIS-Empowered Communication and Localization in Martian Environment}
    \label{sec:ris_in_mars}
    Reflecting on the unique challenges presented by the Martian environment, emerging technologies such as RISs offer promising solutions to enhance communication capabilities. RISs, which rely on passive reflecting surfaces, can potentially mitigate the impact of the Martian surface's distinct electrical properties and enable improved communication in the Martian environment.
    In order to establish the LOS links, different types of RISs need to be intelligently selected and strategically positioned across the diverse geographical formations of the Martian surface. 

    \subsection{Communication Scenarios for RISs in Craters}

    Craters represent one of the most prevalent geographic features on the Martian surface. However, communication between a rover inside a crater and a rover outside the crater encounters certain challenges, particularly due to NLOS links. To address this scenario, a potential solution to enhance the communication link involves deploying RIS at the edge of the crater and optimizing the phases of the reflecting elements of the RIS. The adjustment of the reflecting element phases can be accomplished through two main approaches: CSI-based configuration or localization-based configuration, both of which presuppose the existence of a control link between the transmitter and RIS. CSI can be acquired using semi-passive RISs, as discussed in Section \ref{sec:overview_ris_types}. Alternatively, localization provides another means of determining the optimal configuration for the sub-optimal phase configuration of the RIS. By utilizing the location estimation capability of satellites, the locations of the transmitter and receiver can be determined and sent to the necessary nodes. This extinguishes the need for acquiring CSI and enables RIS to redirect beams to the receiver according to the location information. 
    
\begin{figure}[t]
            % \centering\includegraphics[width=1\linewidth]{figures/Pictureee.png}
            \centering\includegraphics[width=1\linewidth]{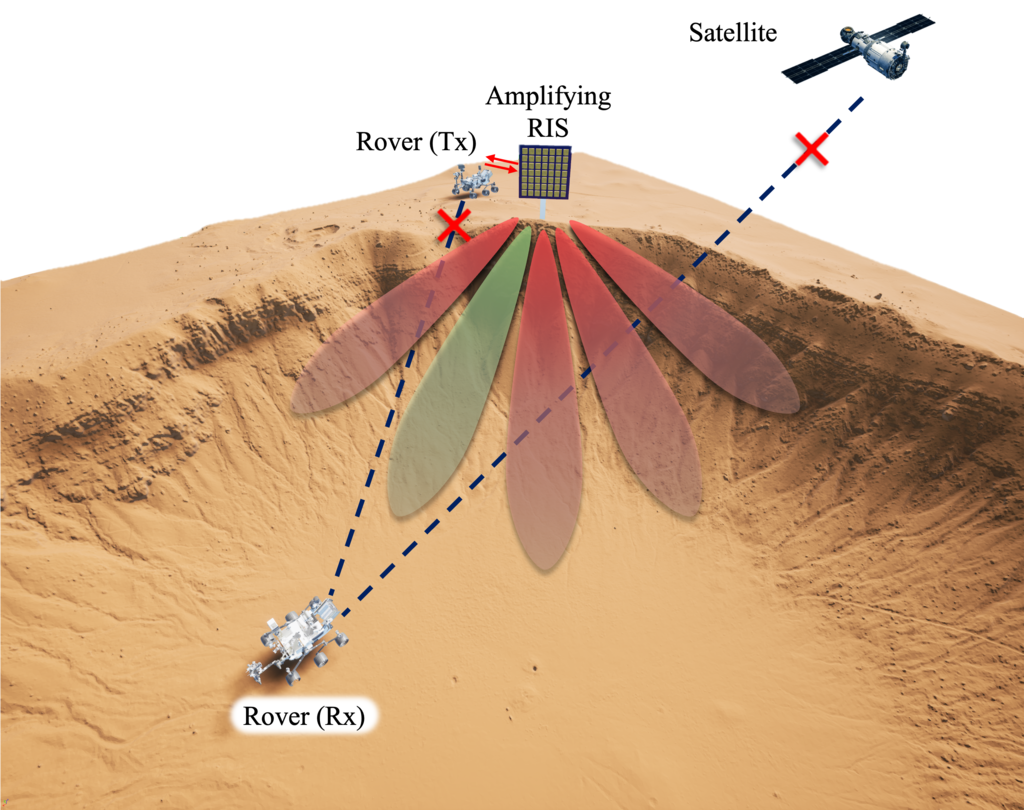}
            %\vspace{10pt}
            \caption{Codebook-based localization system for a rover inside the Mars crater.}
            \label{fig:ntwmodel3}%\vspace*{-0.3cm}
        \end{figure}

    However, there are scenarios where even the satellites are not in LOS with the rovers, such as when the rover is situated in a crater or a canyon while the satellite is near the horizon, as depicted in Fig. \ref{fig:ntwmodel3}. In such cases, it becomes challenging to determine the locations of the transmitter and receiver. RISs can play a crucial role in these scenarios by facilitating the localization process. 
    %By incorporating active elements or sensors on the RIS, the location of the receiver can be determined.
    Nonetheless, a pertinent question is posed by the selection of an appropriate localization algorithm. In this regard, codebook-based strategies in \cite{kayraklik2022indoor} and angular estimation strategy in \cite{Li_user_angle} can be combined. For instance, the scenario depicted in Fig. \ref{fig:ntwmodel3} involves an exploration mission within a crater, where a rover operates, and an amplifying RIS is strategically positioned near the crater's edge to facilitate communication between rovers inside and outside the crater. The transmitter rover, situated outside the crater, seeks to transmit a message to the receiver rover within the crater, utilizing the RIS as a relay medium. Equipped with predefined beams, each associated with different codebooks, the RIS is the focal point of this communication setup \cite{kayraklik2022indoor}.
    The signal received from the transmitter is sequentially radiated by the RIS towards the predefined beams. The signals from each beam are collected by the receiver, and the codebook of the beam corresponding to the largest RSS value is determined \cite{Li_user_angle}.
    Consequently, the RIS is configured to use the determined codebook for the communication between rovers.
    %As a result, the transmitter acquires an understanding of the appropriate beam codebook for utilization within the RIS and transmits its original data via the RISs' selected beam, simultaneously providing the requisite codebook information to the RIS through a control link.
    % While the option exists for the receiver to transfer the channel sounding to the RIS, allowing the RIS to make an optimal beam selection, such an approach results in increased complexity for the RIS. Considering the environmental constraints characteristic of the Martian environment, as discussed earlier, the preference lies in minimizing the intelligence required by the RIS. Consequently, the allocation of the computational workload is primarily to surface operators.

    For the other RIS placement scenarios investigated in the below subsections, the same CSI acquisition or localization approaches are considered while optimizing the phases of the RIS elements.
        
    \subsection{Communication Scenarios for RISs in Plateaus}
    
    In scenarios characterized by LOS conditions, where direct transmission paths exist between the transmitter and receiver, the influence of passive RISs is notably limited compared to situations with NLOS conditions. This is because signals can already reach the receiver without necessitating reflection, and any signal reflected from the RISs suffers from multiplicative path loss. However, the potential contribution of RISs to the signal strength at the receiver in LOS scenarios could be significant if their structures are capable of amplifying the reflected signals. Active RISs, possessing the capacity to strongly reflect by amplifying the incident signals, could thus prove beneficial in LOS scenarios. Notably, in the plateau lands of Mars, where the link between the receiver and transmitter will be predominantly LOS, employing active RIS structures has been deemed preferable to passive RISs, which solely reflect signals \cite{9998527} . Nevertheless, it is important to consider the higher energy consumption associated with active RIS structures compared to passive RISs. Given the scarcity of readily available energy resources on Mars, it is currently not recommended to employ energy-intensive solutions to enhance existing communication systems. Instead, a viable alternative has been proposed in the form of an amplifying RIS structure, which offers similar benefits to other active RISs but with a more energy-efficient approach. Further details about this proposed structure are presented in Section \ref{sec:overview_ris_types}. 
    
    \subsection{Communication Scenarios for RISs in Hills}

    The prevalence of hills on Mars is widely acknowledged, implying that communication links in such environments are prone to NLOS conditions due to these topographic features. For instance, when two rovers are separated by a hill, the obstructed LOS communication drastically degrades the link quality. However, strategically placing an RIS at the summit of this hill can establish a reliable communication link. The selection of the appropriate RIS type can vary depending on the specific scenario. For instance, when the hill lies between the rovers, employing a STAR RIS with appropriate permeability can effectively mitigate the NLOS obstruction. Alternatively, in the same context, the structural advantages of an amplifying RIS enable capturing the signal from one side of the peak, amplifying it, and transmitting the signal further amplified to the other side of the peak, yielding higher energy efficiency compared to active RISs. This scenario is depicted in Fig. \ref{fig:ntwmodel}, illustrating the communication link between Rover $1$ and Rover $3$. Furthermore, when a LOS link already exists between the transmitter and receiver, deploying an amplifying RIS at the top of the hill undoubtedly enhances the communication. This situation is similarly illustrated in Figure \ref{fig:ntwmodel}, dedicated to the link between Rover $1$ and Rover $2$.
    
    \subsection{Communication Scenarios for RISs in Canyons}
    
    \begin{figure}[t]
        % \centering\includegraphics[width=1\linewidth]{figures/Picture2.png}
        \centering\includegraphics[width=1\linewidth]{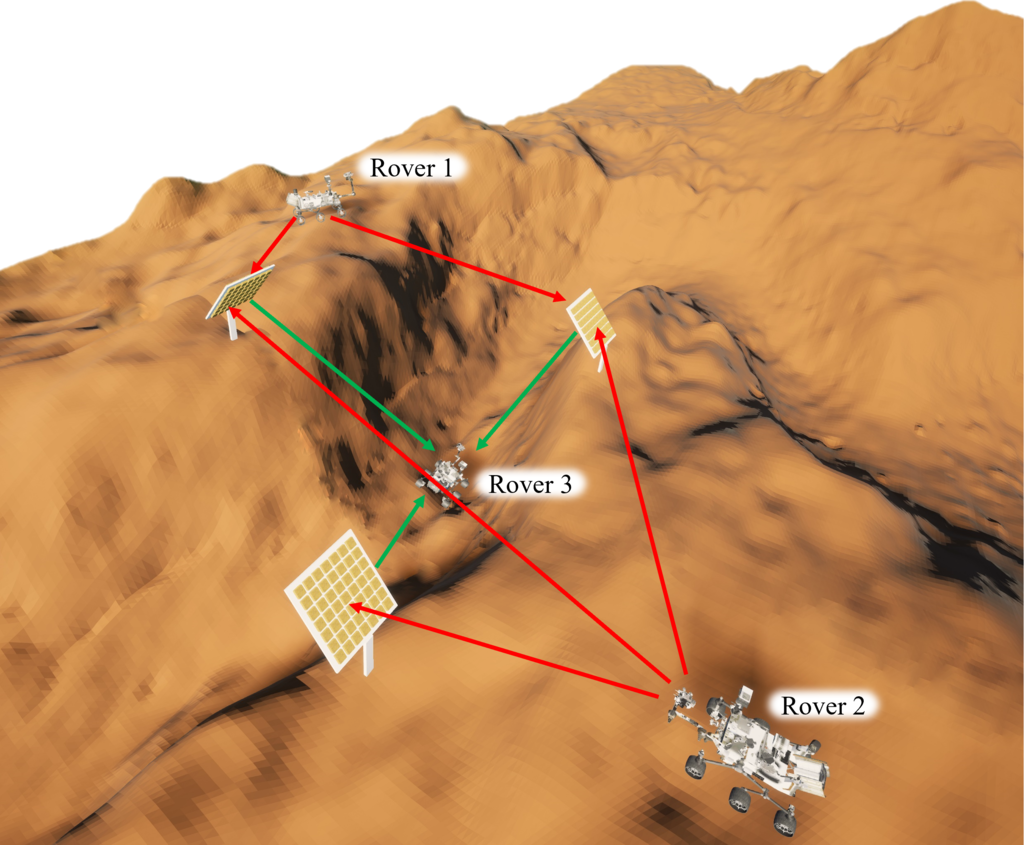}
        %\vspace{10pt}
        \caption{STAR RIS-integrated canyon system.}
        \label{fig:ntwmodel2}%\vspace*{-0.3cm}
    \end{figure}

    Canyons are another most common geological formations on the Martian surface. However, due to NLOS links, establishing communication between rovers inside a canyon and those located outside poses serious challenges. To address these communication limitations, various types of RISs, for example, passive, STAR, and amplifying RISs, can be employed, not only on hills but also within canyons.

    When dealing with the communication links in canyons, a passive RIS can be positioned on the edge of the canyon. This configuration enhances communication quality with the rover inside the canyon by utilizing the reflection. However, this approach may cause additional signal propagation distance. Consequently, placing STAR RISs on the closer side of the canyon, relative to the rover outside, reduces signal travel distance. By introducing dual-sided STAR RISs, positioned on either side of the canyon at distances spanning several hundred meters across, communication support can be extended to both sides of the canyon, enabling the rovers outside to communicate with those inside the canyon through the same RIS. This scenario is presented in Fig. \ref{fig:ntwmodel2}. Additionally, the use of amplifying RISs proves suitable in canyon scenarios. Positioning an amplifying RIS at the closer edge of the canyon to the rover outside ensures reliable communication through amplified signals with the rover positioned inside the canyon.

    Given the curved nature of Martian canyons, rovers within the canyon frequently lack a LOS link with one another. To address this situation, a passive RIS can be placed at the beginning of the canyon's bend, facilitating communication. Additionally, the communication can be reinforced by engaging dual-sided STAR RISs. Moreover, positioning an amplifying RIS perpendicular to the canyon wall at the corner so that it can direct the signal to the other side of the canyon significantly improves the communication link. 
    
    \subsection{Comparison and Performance Evaluation of RIS types and Trade-offs}

    \begin{table}[t]
        \centering
        \caption{Comparison of RIS types under different metrics and scenarios.}
        \label{table_comp}
        \renewcommand{\arraystretch}{1.4}
        \resizebox{8cm}{!}{%
        \begin{tabular}{llccccl}
        \cline{3-6}
         & \multicolumn{1}{l|}{} & \multicolumn{1}{c|}{\begin{tabular}[c]{@{}c@{}}\textbf{Passive}\vspace{-7pt}\\\textbf{RIS}\end{tabular}} & \multicolumn{1}{c|}{\begin{tabular}[c]{@{}c@{}}\textbf{STAR}\vspace{-7pt}\\\textbf{RIS}\end{tabular}} & \multicolumn{1}{c|}{\begin{tabular}[c]{@{}c@{}}\textbf{Amplifying}\vspace{-7pt}\\\textbf{RIS}\end{tabular}} & \multicolumn{1}{c|}{\begin{tabular}[c]{@{}c@{}}\textbf{Active}\vspace{-7pt}\\\textbf{RIS}\end{tabular}} \\ \cline{1-6}
        \multicolumn{1}{|l|}{\multirow{5}{*}{\rotatebox[origin=c]{90}{\textit{\textbf{Metrics}}}}} & \multicolumn{1}{l|}{\begin{tabular}[l]{@{}l@{}}\textbf{Power}\vspace{-7pt}\\\textbf{Consumption}\end{tabular}} & \multicolumn{1}{c|}{Low} & \multicolumn{1}{c|}{Low} & \multicolumn{1}{c|}{Medium} & \multicolumn{1}{c|}{High} \\ \cline{2-6} 
        \multicolumn{1}{|l|}{} & \multicolumn{1}{l|}{\textbf{Complexity}} & \multicolumn{1}{c|}{Low} & \multicolumn{1}{c|}{Low} & \multicolumn{1}{c|}{Medium} & \multicolumn{1}{c|}{High}\\ \cline{2-6}
        \multicolumn{1}{|l|}{} & \multicolumn{1}{l|}{\textbf{Performance}} & \multicolumn{1}{c|}{Low} & \multicolumn{1}{c|}{Low} & \multicolumn{1}{c|}{High} & \multicolumn{1}{c|}{High}\\ \cline{2-6}
        \multicolumn{1}{|l|}{} & \multicolumn{1}{l|}{\textbf{Cost}} & \multicolumn{1}{c|}{Low} & \multicolumn{1}{c|}{Low} & \multicolumn{1}{c|}{Medium} & \multicolumn{1}{c|}{High}\\ \cline{2-6}
        \multicolumn{1}{|l|}{} & \multicolumn{1}{l|}{\begin{tabular}[l]{@{}l@{}}\textbf{Energy}\vspace{-7pt}\\\textbf{Efficiency}\end{tabular}} & \multicolumn{1}{c|}{Low} & \multicolumn{1}{c|}{Low} & \multicolumn{1}{c|}{High} & \multicolumn{1}{c|}{Medium}\\ \cline{1-6}
        \multicolumn{1}{|l|}{\multirow{4}{*}{\rotatebox[origin=c]{90}{\textit{\textbf{Scenarios}}}}} & \multicolumn{1}{l|}{\textbf{Canyon}} & \multicolumn{1}{c|}{\cmark} & \multicolumn{1}{c|}{\cmark} & \multicolumn{1}{c|}{\cmark} & \multicolumn{1}{c|}{\cmark} \\ \cline{2-6}
        \multicolumn{1}{|l|}{} & \multicolumn{1}{l|}{\textbf{Crater}} & \multicolumn{1}{c|}{\cmark} & \multicolumn{1}{c|}{\cmark} & \multicolumn{1}{c|}{\cmark} & \multicolumn{1}{c|}{\cmark} \\ \cline{2-6}
        \multicolumn{1}{|l|}{} & \multicolumn{1}{l|}{\textbf{Mountain}} & \multicolumn{1}{c|}{\xmark} & \multicolumn{1}{c|}{\xmark} & \multicolumn{1}{c|}{\cmark} & \multicolumn{1}{c|}{\cmark} \\ \cline{2-6}
        \multicolumn{1}{|l|}{} & \multicolumn{1}{l|}{\textbf{Plateau}} & \multicolumn{1}{c|}{\xmark} & \multicolumn{1}{c|}{\xmark} & \multicolumn{1}{c|}{\cmark} & \multicolumn{1}{c|}{\cmark} \\ \cline{1-6}
        \end{tabular}%
        }
    \end{table}
    As elucidated in the preceding sections, different types of RISs possess distinct advantages and disadvantages in various scenarios. The comparison of four different RIS types with various metrics is presented in Table \ref{table_comp}, along with an assessment of their suitability for deployment in different geographical formations on the Martian surface \cite{ebasarwctris}, \cite{9998527}, \cite{9758764}, \cite{9437234}. However, it is essential to underscore that despite active RISs being viable in certain cases, their utilization in Martian communications is deemed unsuitable due to higher energy consumption, as discussed earlier.

    Table \ref{table_comp} highlights the low power consumption of passive and STAR RISs, attributable to their lack of active elements. Conversely, active RISs exhibit high power consumption due to the inclusion of multiple amplifiers. Specifically, the amplifying RIS, with a single power amplifier, falls within an intermediate range of power dissipation. Nonetheless, authors of \cite{9758764} have demonstrated that, unless employing very high gain and output power values, the amplifying RIS's power consumption is negligibly higher than that of passive RISs. From a hardware perspective, the amplifying RIS remains relatively simple, as it essentially consists of two RIS layers with an amplifier placed in between. This simplicity implies that additional requirements for power budgeting and thermal management are minimal compared to fully active RIS architectures. Moreover, when considering extreme environments such as Mars, where large temperature fluctuations, radiation, and dust storms are present, the robustness of the amplifying RIS is unlikely to be a concern. If highly sophisticated and power-demanding systems such as Mars rovers can reliably operate under such conditions, the modest hardware of an amplifying RIS would similarly function without significant degradation. Therefore, while practical considerations such as thermal stability must be acknowledged, the amplifying RIS can be regarded as a robust and energy-efficient solution even in challenging environments. Consequently, the complexity and cost of the amplifying RIS are moderate, while those of passive and STAR RISs are low, and those of active RISs are high.

    In terms of performance, active RISs and amplifying RISs exhibit high efficacy as they amplify the incoming signal before transmitting it to the receiver. On the other hand, STAR and passive RISs do not involve any signal amplification operation, resulting in comparatively lower performance than their active and amplifying counterparts.

    In general, the deployment of an amplifying RIS is warranted in scenarios where there is a considerable distance between the RIS and both the transmitter and receiver or when a LOS link is established between the transmitter and receiver. Instances of such scenarios are observed in geographical formations like high hills or plateaus, where the impact of using passive RIS or STAR RIS is limited. Moreover, passive RIS or STAR RIS can be effectively applied in structures such as canyons and craters. However, even in these scenarios, the integration of amplifying RIS will lead to a further enhancement in the RSS.

    \subsection{Control signaling and RIS reconfiguration under discontinuous Martian links.}

    We adopt a low-overhead control plane that decouples illumination from channel estimation. The controller (onboard a rover, a nearby lander, or an orbiter when in view) maintains a small codebook of RIS phase profiles that map to sectors or wide illumination regions covering the operational area. At the beginning of a block, the rover’s position (or direction-of-departure in far field) is estimated using the available links; the RIS phase profile is then selected accordingly. During data transmission, the mobile node monitors the average SNR; when this metric falls below a predefined threshold, it triggers a reconfiguration cycle that updates position and switches the RIS profile. In this architecture, the reconfiguration rate is governed by mobility and the width of the illumination, not by the short channel coherence time. Analyses of such decoupled designs show that the required overhead is negligible compared to CSI-centric schemes and that widening the illumination reduces the need for frequent updates at modest SNR targets \cite{9673796}. The control plane is described as operating on slow geometry with event driven updates, so precise global time alignment is not required; reconfiguration is triggered by a long term SNR threshold and the update period is seconds to minutes. The localization layer may use received signal strength or angle information and a short codebook sweep at the RIS, which avoids dependence on time of arrival when clocks are not tightly synchronized. This control plane fits Mars well: it tolerates intermittent backhaul, minimizes signaling exchanges, and can operate with pre-stored phase profiles when links are temporarily unavailable. For completeness, we also note that full-coverage illumination is a special case that eliminates reconfiguration at the expense of higher transmit power and larger surfaces; this option may be attractive for static users or safety-critical beacons. Lastly, the required localization accuracy is sub-meter and the update period is on the order of seconds to minutes, which can already be achieved by rover visual–inertial odometry without global navigation satellite system. Occasional orbiter passes are used only to bound drift. The control loop operates continuously on onboard estimates and transmits only compact codebook indices when a reconfiguration is triggered.
    
    \section{Simulation Results of RISs on Martian Surface}
    \begin{figure}[t]
            % \centering\includegraphics[width=1\linewidth]{figures/Pictureee.png}
            \centering\includegraphics[width=0.9\linewidth]{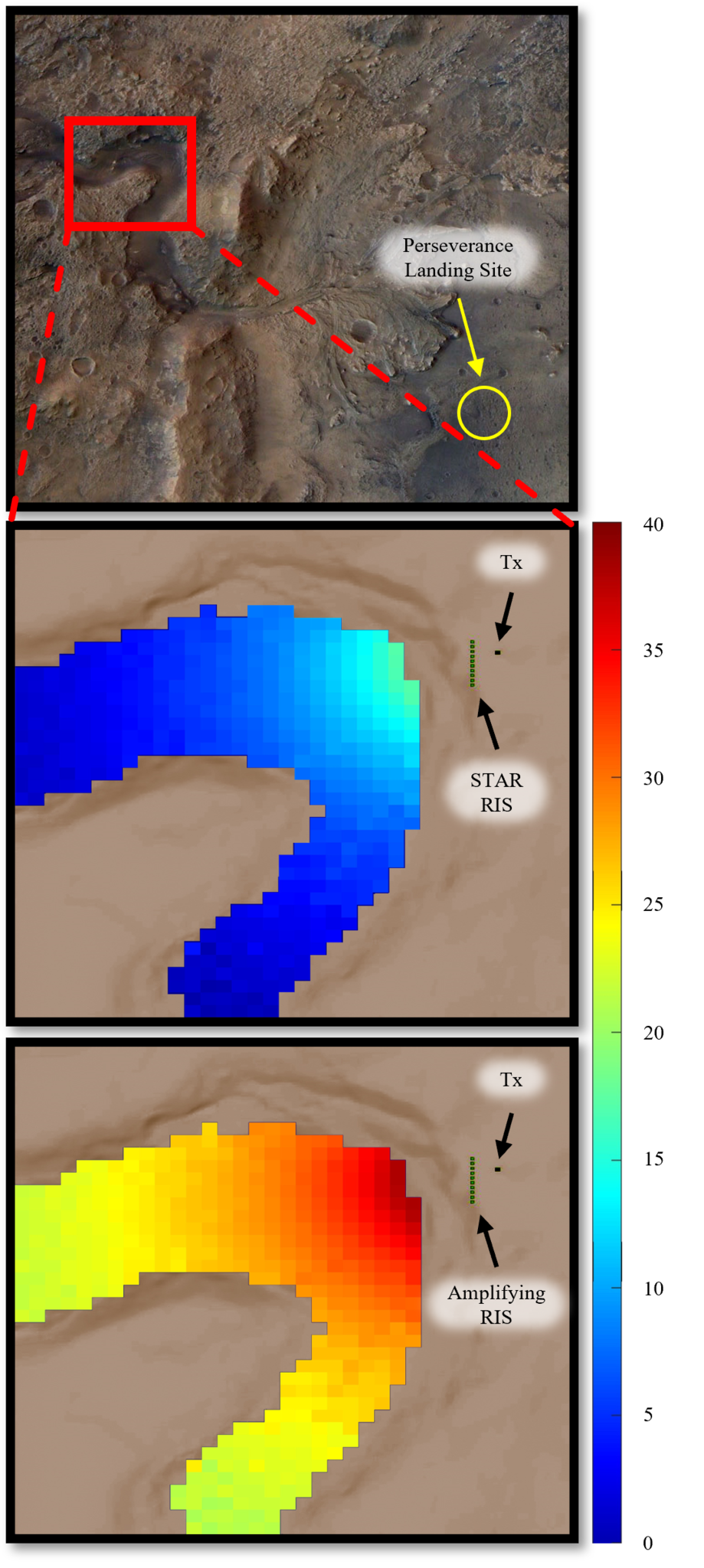}
            %\vspace{10pt}
            \caption{Heatmap of the received SNR [dB] values in Neretva Vallis for STAR RIS and amplifying RIS.}
            \label{fig:coverage}%\vspace*{-0.3cm}
        \end{figure}
    % Heatmap anlatımı
    Computer simulations were conducted in the Martian environment using a ray tracing-based channel modeling software called Wireless Insite \cite{remcom_wireless_insite}. The purpose is to compare the performance of STAR RIS and amplifying RIS in irregular terrain scenarios. The simulations focused on the Jezero crater, which serves as the landing site for the Perseverance rover. This crater, measuring \SI{45}{\kilo\meter} wide, was once submerged and contained an ancient river delta. The simulations were centered around a specific area called Neretva Vallis, believed to be the location where water entered the crater.
    The electrical properties of the Martian terrain set to the conductivity of \SI[per-mode = symbol]{e-8}{\siemens\per\meter} and permittivity of \SI[per-mode = symbol]{4}{\farad\per\meter}. The electrical parameters used in the simulator reflect a dry, basaltic Martian surface with very low loss. This choice is conservative for our frequency band because it reduces natural reflections and therefore does not exaggerate the benefit of RIS. At sub-$6$ \SI{}{\GHz}, the line of sight remains the dominant component, and reasonable variations in the ground parameters only cause small changes in the reflection coefficient and in the preferred RIS orientation. We add that a brief in situ calibration can refine the RIS phases at deployment time, so the system operation does not depend on exact prior knowledge of the regolith properties.
    The simulations were carried out at a frequency of \SI{5}{\GHz}. We believe that if we stay at this frequency range, the beam‑squint effect will be negligible. However, it needs to be explored for mmWave/Terahertz scenarios, where phase‑shifters present some limitations causing squint. Atmospheric conditions, such as temperature (\num{-63}\si{\celsius}), pressure (\num{6.1}\si{\milli\bar}), and humidity (\num{20}$\si{\percent}$), were also taken into account. During the simulations, practical parameters are considered for the rovers, encompassing a transmit power of \SI{10}{dBm}, receiver antenna gain of \SI{20}{dBi}, transmit antenna gain of \SI{20}{dBi}, and low-noise amplifier (LNA) gain of \SI{10}{dB}. To prevent excessive power consumption, a conservative value of \SI{10}{dB} is selected as the gain for the power amplifier utilized in the amplifying RIS. Increasing the amplifier gain to \SI{15}{dB} or \SI{20}{dB} further enhances the received SNR, though with diminishing returns \cite{9758764}. However, this improvement comes at the cost of a considerable increase in power consumption compared to the passive design, which would reduce the overall energy efficiency and complicate thermal management. Therefore, \SI{10}{dB} represents a balanced compromise, providing meaningful SNR gains while keeping the power budget within practical limits. Additionally, an amplifier noise figure of \SI{5}{dB} is chosen. The phases of the RISs are configured to eliminate the phase shifts introduced by the Tx-RIS and RIS-Rx channels. The magnitude of the reflection or transmission coefficient for the reflecting elements is assumed to be \num{1}. We note that when the ideal element magnitude and perfect phase alignment are relaxed to include realistic loss and phase errors, together with moderate variability in regolith permittivity and conductivity, the absolute SNR may decrease by a few decibels while the relative ranking between STAR RIS and amplifying RIS observed in our scenarios remains unchanged. Furthermore, the noise power at the receiver is assumed to be \SI{-100}{dBm}. The performance of the communication link between the transmitter, the RIS, and the receiver is evaluated by generating a heatmap based on the SNR values for each receiver grid.
    Fig. \ref{fig:coverage} illustrates the significant disparity in signal quality between STAR RIS and amplifying RIS.
    Upon examining the SNR ranges of the receivers, it becomes evident that STAR RIS yields SNRs ranging from \SI{0}{dB} to \SI{20}{dB}, whereas the amplifying RIS provides SNRs between \SI{20}{dB} and \SI{40}{dB}. Thus, a notable \SI{20}{dB} difference in SNR is observed between the two types of RISs. Of this \SI{20}{dB} difference, \SI{10}{dB} can be attributed to the amplifier gain, while the remaining difference appears to be due to the beamforming gain. In this context, it has been observed that the incorporation of amplifying RISs can significantly enhance both coverage and signal quality.

    \section{Future Directions} \label{sec:future_directions}

    The door is open for further research on communications for the planet Mars. Firstly, more sophisticated simulation environments, which takes aspects such as low-density atmosphere, presence of carbon dioxide, reflection coefficient, dust storms and radiation from the Sun into account in each iteration need to be developed to enable more accurate analysis of Martian communication scenarios.

    \subsection*{RIS-aided beamforming}
    There are also promising research fields in the area of RIS communication that can further improve and diversify their use cases in Mars communication. For example, \cite{9020088} developed one of the first RIS prototypes to demonstrate real-time beamforming gains in wireless communications. The authors confirm the feasibility of RIS-aided beamforming with much lower power consumption than a conventional phased array.
    \cite{9454375} examines the utilization of an RIS to enhance radar sensing capabilities. It presents a sensing architecture in which a radar acquires an additional echo path through the RIS, therefore enabling the radar to detect a target that would otherwise be obscured or outside its LoS. The RIS functions as an adjustable mirror, reflecting radar probing signals both toward and away from the target. Using this method, the authors compute detection performance enhancements and demonstrate that RIS-aided radar can improve target detection probability in challenging scenarios (e.g., around corners or beyond obstacles) when compared to a standalone radar. 
    Beamforming on Mars can be combined with local sensing, as the latency to Earth makes distant beam training difficult. A viable option is to include semi-passive detectors in each RIS tile that capture a small amount of the incident wave, allowing for real-time angle-of-arrival estimation and coarse channel state determination without the need for an energy-consuming control link. By reusing the same samples for communication and sensing, the surface can quickly guide its beams toward a moving rover while also providing bearing information, which allows self-localization inside craters or canyons where satellite GPS is blocked.
        
    \subsection*{Modulation}
    Future Martian sensor networks can benefit from novel modulation schemes that use the RIS as the single RF modulator, avoiding the mass and energy expenses of traditional transmit chains. \cite{10838608} have recently shown a direct-data backscatter architecture in which the RIS imposes digital symbols on an incident carrier while steering the reflected beam toward the receiver. Translating this approach to Mars involves creating reflection-modulation constellations and adaptive beam-steering codes that is capable of handling the planet's delay spreads, weak multipath, and Doppler shifts. Such RIS-aided backscatter would allow low power IoT nodes, such as subterranean humidity probes or dust-storm sensors, to feed data onto relay reflections without the need for active radios, preserving scarce energy and spectrum while expanding coverage into canyons.
    
    \subsection*{Wireless Absorption and Power Transfer}
    Absorption and power transfer techniques can also be used to extend the functionality of RISs in Martian environment. RISs may reflect or absorb incoming radio waves by tunning their tiny on-board circuits. Experiments by \cite{9092996} reveal that a surface built in this way may absorb nearly all of the energy that reaches it, which is beneficial for preventing interference or harvesting stray power for sensors. \cite{10547218} show that a single RIS can control part of a beam toward a device that needs to be recharged while also steering a separate beam to transmit data to another user. On Mars, the same dual-purpose panels may gather downlink energy to charge dust-storm sensors or redirect it into a tight communication beam that avoids canyon walls. In the planet's energy-constrained environment, combining absorption, power transfer, and signal relaying into a single RIS would reduce payloads and increase network life.

\section{CONCLUSION} \label{sec:Conclusions}
    The article has investigated the use of RIS-assisted localization, presenting a general framework for estimating user positions in challenging Martian conditions. As space exploration, particularly missions to Mars becomes more important, it is crucial to have reliable and robust communication systems in place. The RIS technology offers promising solutions to improve communication coverage and links in various Martian scenarios, such as canyons, craters, mountains, and plateaus.
    The findings of this article reveal significant differences in performance between different types of RISs, with amplifying RISs showing higher SNRs compared to STAR RISs. We further emphasize the importance of considering different types of RIS and their optimal placement in various Martian terrains.    
    As we aspire to colonize Mars and undertake future missions, incorporating the RIS technology can ensure more reliable and robust communication links. Furthermore, this article opens up promising research directions for integrating RISs into deep space communication and improving interplanetary communication networks. By embracing the RIS technology, future Mars missions can benefit from improved communication links, enabling us to further explore and understand the Red Planet.

	\ifCLASSOPTIONcaptionsoff
	\newpage
	\fi

    \bibliographystyle{IEEEtran}
	\bibliography{mars_ref}

\end{document}